\documentclass[14pt]{article}
\title{Quantum Mechanics in Phase Space.}
\author{Ali Nassimi \\Department of Chemistry,
University of Toronto,\\Toronto, ON, M5S 3H6,
Canada\\ali.nassimi@utoronto.ca}
\usepackage{setspace}

\date{}
\begin{document}
\maketitle
\bibliographystyle{unsrt}
\tableofcontents
\begin{abstract}
The basics of the Wigner
formulation of Quantum-Mechanics and few related interpretational issues are presented in a simple language.
 This formulation has extensive applications in Quantum Optics and in Mixed Quantum-Classical formulations.
\end{abstract}
\section{Introduction}
In classical mechanics, a system is completely specified by the
position and momentum of its particles. It is common to build a
6\emph{N} dimensional space, where \emph{N} is the number of
particle constituting the system. This space is called the phase
space and contains an axe for each of the coordinate and momentum
of the system. The system is completely specified by a point in
this space. Usually, when we have a lack of knowledge regarding
the state of the system, we try to find a probability distribution
for the system in its phase space. This is the case in classical
statistical mechanics, where $f(p,q)$ ($p$ is a 3$N$ dimensional
vector representing the momentum of all particles, and $q$ is a
3$N$ dimensional vector representing the coordinate of all
particles) is the probability of finding the system with
coordinate between $q$ and $q+dq$ and momenta between $p$ and
$p+dp$. The predictions of quantum mechanics are similar to that
of classical statistical mechanics in the sense that they are
statistical in nature. Thus, the fundamental question arise
whether quantum processes can be described as an average over
uniquely determined processes or not. And the observability of
these processes if the answer is yes \cite{groenewold}. Naturally
the place to look for such processes is the phase space and if the
system is undergoing a process in the phase space it must be
represented by a point in the phase space. Regardless of the tough
problem mentioned, it seems natural for a scientist to seek a
function similar to distribution functions in quantum mechanics.
But, in the framework of Orthodox and Copenhagen interpretations
of quantum mechanics this is impossible.
\section{Wigner distribution function} The first example of such a function in quantum
mechanics was suggested by Wigner \cite{Wigner32}. He mentioned
the canonical ensemble probability distribution in classical
statistical mechanics for a system having momenta between $p$
and $p+dp$ and coordinate between $q$ and $q+dq$, which is
$e^{-\beta \epsilon}$, where $\beta$ is the reciprocal of the
temperate, $T$, and $\epsilon$ is the sum of the kinetic and the
potential energy. He mentioned that in quantum mechanics, we
cannot simultaneously have momentum and position so we cannot have
such an expression. But even if we consider the coordinate alone,
where the classical expression for probability is $e^{-\beta V}$,
where V is the potential energy of the system, the classical
expression is not valid for quantum systems, because when $\beta
\rightarrow \infty$ there is no reason for that expression to be
equal to $|\psi_0(x_1,...,x_n)|^2$ (the ground state wave function
is not even always known). Although, the statistical mechanics of
quantum systems is given by the von Newman formula, i.e.,
$<Q>=Tr(Q e^{- \beta H})$, where $Q$ is the operator corresponding
to the quantity under consideration, $H$ is the Hamiltonian of the
system and $<>$ denote the expectation value. Since, it was not
easy to use the von Newman formula for evaluating the expectation
values, Wigner suggests to build the following expression
\begin{equation}\label{wigner1}
P(q,p)=(\frac{1}{2\hbar \pi})^n \int dy \psi(q-y/2)
\psi(q+y/2)^* e^{ipy/\hbar},
\end{equation}
and call it the probability function, here n is the dimension of the space. Thorough out this paper consider the
limits of the integrals from $-\infty$ to $\infty$, unless
otherwise is explicitly stated. Unfortunately, Wigner never
mentioned how he come up with this recipe, he just mention: "This expression was
found by L. Szilard and the present author some years ago for
another purpose." By introducing the inverse Fourier transform of
$\psi$, i.e., $\psi(q)=(\frac{1}{2\pi \hbar})^{\frac{n}{2}}\int dp
e^{\frac{i}{\hbar}pq} \psi(p)$ in the above relation we could get
\begin{equation} \label{wigner}
\int dy \int dp' \int dp''\psi(p')^* \psi(p'')
e^{\frac{i}{\hbar}[-p'(q+y/2)+p''(q-y/2)+py]},
\end{equation}
by performing the integral over $y$ we could get $(2 \pi \hbar)^n
\delta(p-\frac{p''+p'}{2})$, then we can perform the integral over
$p''$ and perform the change of variables $p-p' \rightarrow
-y/2$to get
\begin{equation}\label{wigner2}
P(q,p)=(\frac{1}{2\hbar \pi})^n \int dy \psi(p-y/2)
\psi(p+y/2)^* e^{iqy/\hbar}.
\end{equation}
This relation is completely equivalent with the relation
(\ref{wigner1}) and shows the symmetry of the Wigner functions
with respect to $q$ and $p$. The phase space function
corresponding to an operator $A$ is defined thorough
\begin{equation}\label{wigner3} A(q,p)=\int
dy e^{ipy/\hbar}<q-\frac{y}{2}|\hat{A}|q+\frac{y}{2}>.
\end{equation}
\section{Proposals for getting the Wigner function} 
Stenholm presents a derivation for the Wigner function
\cite{stenholm}. All the information extractable from the quantum
theory is contained in the matrix elements
\begin{equation}\label{derivation1}
<x_1|\hat{\rho}|x_2>=\psi(x_1)\psi(x_2)^*.
\end{equation}
 We can bring the density matrix into momentum
representation and write
\begin{equation}\label{derivation2}
<p_1|\hat{\rho}|p_2>=\frac{1}{2\pi \hbar}\int\int dx_1 dx_2
exp[-i(p_1x_1-p_2x_2)/\hbar]<x_1|\hat{\rho}|x_2>.
\end{equation}
Similar to a two body problem in mechanics, we can define new
variables as $R=\frac{x_1+x_2}{2}$ and $r=x_1-x_2$, and a similar
change of variables in the momentum representation, i.e.,
$P=\frac{p_1+p_2}{2}$ and $p=p_1-p_2$. It is simple to show that
\begin{equation}\label{derivation3} p_1x_1-p_2x_2=Pr+pR.
\end{equation} By substituting (\ref{derivation3}) into
(\ref{derivation2}) and changing the variables, we could get
\begin{equation}\label{derivation4}
<P+\frac{p}{2}|\rho|P-\frac{p}{2}>=\frac{-1}{2\pi \hbar}\int\int
dr dR exp[-i(Pr+pR)/\hbar]<R+\frac{r}{2}|\rho|R-\frac{r}{2}>.
\end{equation}
The above relation is just the Fourier
transform of $\rho(R,r)$, where by analogy to the two particle
problem, we can call $R$ the center of mass coordinate, and $r$
the relative coordinate. Because, we are interested to get a
function containing both momentum and coordinate, we could either
drop the Fourier transformation on relative coordinate to get the
Wigner function, or drop the Fourier transformation on the center
of mass coordinate to get the Shirley \cite{shirley} function.

Groot has presented another equivalent method for deriving the
Wigner function \cite{groot}. By inserting unity operators, we get
\begin{equation}\label{derivation5}
\hat{A} = \int dp'dp''dq'dq''|q''><q''|p''><p''|A|p'><p'|q'><q'|.
\end{equation}
Then, we can introduce the new variables $p'=p-u/2$, $p''=p+u/2$,
$q'=q-v/2$ and $q''=q+v/2$, where the Jacobian is equal to one,
and use the relation $<q|p>=h^{-n/2}e^{\frac{i}{\hbar}p.q}$ to get
\begin{equation}\label{derivation6}
 \begin{array}[c]{c}
\hat{A}=h^{-n} \int
dpdqdudv|q+v/2>e^{\frac{i}{\hbar}(p+u/2)(q+v/2)}<p+u/2|A|p-u/2> \\
\\e^{\frac{-i}{\hbar}(p-u/2)(q-v/2)}<q-v/2|.
\end{array}
\end{equation}
This relation simplifies to
\begin{equation}\label{derivation7}
\hat{A}= h^{-n} \int dpdqdudv
|q+v/2><p+u/2|A|p-u/2><q-v/2|e^{\frac{i}{\hbar}(qu+pv)}.
\end{equation}
By defining the $\hat{A}$ dependent function
\begin{equation} \label{derivation8}
a(p,q)=\int du<p+u/2|A|p-u/2>e^{\frac{i}{\hbar}qu},
\end{equation}
and the $\hat{A}$ independent operator
\begin{equation} \label{derivation9}
\hat{\Delta}(p,q)=\int dv |q+v/2><q-v/2|e^{\frac{i}{\hbar}pv}.
\end{equation}
We have
\begin{equation} \label{derivation10}
A=h^{-n}\int dpdq a(p,q) \hat{\Delta}(p,q).
\end{equation}
It is clear that $a(p,q)$ is the Wigner function corresponding to
the operator $\hat{A}$. This is a natural way one can lead to the
definition of the Wigner function.

\section{Weyl operator}
Before this work by Wigner, Weyl \cite{weyl} had proposed a
method to construct an operator $\hat{A}$ corresponding to the
phase space function $A(q,p)$. First we define
\begin{equation}\label{weyl1}
\alpha(\sigma,\tau)= \left(\frac{1}{2 \pi \hbar}\right)^n \int dq
\int dp e^{-i(\sigma q+\tau p)/\hbar} A(q,p)
\end{equation}
and then,
\begin{equation}\label{weyl2}
\hat{A}(\hat{q},\hat{p})=\int d\sigma \int d\tau
\alpha(\sigma,\tau)e^{i(\sigma \hat{q}+\tau \hat{p})/\hbar}
\end{equation}
Wigner's recipe is exactly the inverse of the Weyl's. If this is a
suitable correspondence between $A(p,q)$ and $\hat{A}$, so we must
be able to get the correct expectation value for $\hat{A}$ by use
of $A(p,q)$, i.e.,
\begin{equation}\label{weyl3}
<\psi|\hat{A}|\psi>=\int dq \int dp P(q,p)A(q,p).
\end{equation}
Before proving this equality, I should mention a lemma.

lemma 1: By using the Baker-Hausdorff lemma, we can prove that
\begin{equation}\label{lemma1}
e^{\hat{A}+\hat{B}}=e^{\hat{A}}e^{\hat{B}}e^{\frac{-1}{2}[A,B]},
\end{equation}
 which yields to
\begin{equation}\label{lemma2}
e^{\frac{i}{\hbar}(\sigma \hat{q}+\tau
\hat{p})}=e^{\frac{i}{\hbar}\sigma \hat{q}}e^{\frac{i}{\hbar}\tau
\hat{p}}e^{i\sigma \tau /2\hbar}.
\end{equation}
By substituting $A(p,q)$ from (\ref{weyl1}) and $\hat{A}$ from
(\ref{weyl2}) into (\ref{weyl3}), we get
 \begin{equation}\label{weyl4}
\begin{array}[c]{c} \int d\sigma \int d\tau \alpha(\sigma,\tau)
<\psi|e^{i(\sigma \hat{q}+\tau \hat{p})/\hbar}|\psi>\\ \\ = \int
d\sigma \int d\tau \int dq \int dp P(q,p)e^{i(\sigma q+\tau
p)/\hbar} \alpha(\sigma,\tau),
\end{array}
\end{equation}
which easily simplifies to
\begin{equation}\label{weyl5}
\begin{array}[c]{c}
<\psi|e^{i(\sigma \hat{q}+\tau \hat{p})/\hbar}|\psi>=\int dq \int
dp P(q,p)e^{i(\sigma q+\tau p)/\hbar} \\ \\= (2\pi \hbar)^{-n}
\int dy \int dq \int dp \psi(q+y)^* \psi(q-y) e^{i(2py+\sigma
q+\tau p)/\hbar}.
\end{array}
\end{equation}
The integral over $p$ gives $(2 \pi \hbar)^n \delta(2y+\tau)$,
which allow us to perform the integral over $y$ in order to get
for the right hand side
\begin{equation}\label{weyl6}
\int dq \psi(q+\tau/2)^* \psi(q-\tau/2) e^{i(\sigma q)/\hbar}.
\end{equation}
According to the lemma 1, the left hand side is
\begin{equation}\label{weyl7}
e^{i\sigma \tau/2\hbar}<\psi|e^{\frac{i}{\hbar}\sigma
\hat{q}}e^{\frac{i}{\hbar}\tau \hat{p}}|\psi>.
\end{equation}
Because $p$ is the generator of translation (\ref{weyl7}) is equal
to
\begin{equation}\label{weyl8}
 \int dx e^{\frac{i}{\hbar}(\sigma
x+\sigma \tau /2)} \psi(x)^* \psi(x+\tau).
\end{equation}
By imposing the change of variable $x\rightarrow q-\tau /2$, we
get the relation (\ref{weyl6}), so Q.E.D.

\section{Properties of the Wigner distribution} \label{Properties}
A number of properties have been mentioned for this function
\cite{hillery}

(i) Since $P(q,p)$ should be real, it should be corresponding to a
Hermitian operator, i.e.,
\begin{equation}\label{property1}
P(q,p)=<\psi|M(q,p)|\psi>,
\end{equation} where $M=M^\dag$, i.e., Hermitian.

(ii)
\begin{equation}\label{property2a}
\begin{array}[c]{c}
\int dp P(q,p)=(\frac{1}{2\pi \hbar})^n\int dp \int dy <q-y/2|\rho|q+y/2> e^{ipy/\hbar} \\ \\
=\int dy \delta(y) <q-y/2|\rho|q+y/2> =|\psi(q)|^2=<q|\rho|q>.
\end{array}
\end{equation}
\begin{equation}\label{property2b}
\int dq P(q,p)=|\psi(p)|^2=<p|\rho|p>
\end{equation}
 \begin{equation}\label{property2c}
\int dq \int dp P(q,p)=Tr(\rho)=1
\end{equation}
 Derivation of the second and the
third one are similar to that of the first one.

(iii) Translation of $P(q,p)$ in the momentum and coordinate
spaces occur in accordance with the translation of the wave
function, i.e., if $\psi(q) \rightarrow \psi(q+a)$ then $P(q,p).
\rightarrow P(q+a,p)$, and if $\psi(q) \rightarrow
e^{ip'q/\hbar}\psi(q)$ then $P(q,p) \rightarrow P(q,p-p')$ (iv)
$P(q,p)$ should change the same way as $\psi$ in space reflections
and time inversions , i.e., if $\psi(q) \rightarrow \psi(-q)$,
then $P(q,p) \rightarrow P(-q,-p)$ and, if $\psi(q) \rightarrow
\psi(q)^*$ then $P(q,p) \rightarrow P(q,-p)$ 

(v) When the third
and all higher order derivatives of the potential are zero we get
the classical equations of motion (the Liouville equation). (This
will be shown in the section \emph{Dynamics}.)

(vi)
\begin{equation}\label{property3}
|<\psi(q)|\phi(q)>|^2=2\pi \hbar \int dq \int dp P_\psi(q,p)
P_\phi(q,p)
\end{equation}

(vii)
\begin{equation}\label{property4}
\int dq \int dp A(q,p) B(q,p)=2\pi\hbar \rm{Tr}(\hat{A}\hat{B}),
\end{equation}
where $A(q,p)$ is the classical function corresponding to the
quantum operator $A$. Using the property (ii) it can be easily
shown that if $h(q,p)=f(q)+g(p)$ then we can get the expectation
value of $h$ by $\int \int dp dq P(q,p)[f+g]$

\section{The Product of two Operators}
Groenewold in a fundamental work presented some foundational issues of
quantum mechanics. He depicts
the physical properties corresponding to the quantum mechanical
operators $\hat{A}$ and $\hat{B}$ with $a$ and $b$. He used the
von Newman's assumptions, i.e., (I) if $a$ corresponds to
$\hat{A}$ and $b$ corresponds to $\hat{B}$ then $a+b$ corresponds
to $\hat{A}+\hat{B}$, and (II) if $a$ corresponds to $\hat{A}$
then $f(a)$ corresponds to $f(\hat{A})$. He shows that
such symbols constitute two isomorphic groups. Thus, if $\hat{A}$
and $\hat{B}$ do not commute then $a$ and $b$ should not commute.
It can be shown that by assuming $a$ and $b$ as commuting
observables we get into contradiction with assumptions (I) and
(II). Therefore, a quantum system can not possess two physical
properties corresponding to two non-commuting operators, and there
is no reason to introduce different notation for operator and
physical property. In that paper, he shows
\begin{equation}\label{product1}
\hat{A}\hat{B}=\hat{F} \rightarrow F(q,p)=A(q,p)e^{(\hbar \Lambda
/2i)}B(q,p)=B(q,p)e^{-(\hbar \Lambda /2i)}A(q,p),
\end{equation}
where
\begin{equation}\label{product2}
\Lambda=\frac{\overleftarrow{\partial}}{\partial
p}\frac{\overrightarrow \partial}{\partial
q}-\frac{\overleftarrow{\partial}}{\partial
q}\frac{\overrightarrow
\partial}{\partial p},
\end{equation}
is the negative of the Poisson bracket. Note that there is a dot product between the differentiation
toward right and differentiation toward left. By taking the matrix elements
of (\ref{weyl2}), we get
\begin{equation}\label{product3}
<q''|\hat{A}|q'> = \int d \sigma \int d \tau
\alpha(\sigma,\tau)<q''|e^{i(\sigma \hat{q}+\tau
\hat{p})/\hbar}|q'>.
\end{equation}
and by using the lemma 1, we can get
\begin{equation}\label{product4}
\begin{array}[c]{c}
<q''|\hat{A}|q'>= \int d \sigma d\tau
\alpha(\sigma,\tau)e^{i \sigma \tau /2\hbar} e^{i\sigma
(q'-\tau)/\hbar} \delta(q'-\tau-q'') \\ \\ = \int d\sigma \alpha(
\sigma , q'-q'') e^{i\sigma (q'+q'')/2\hbar}.
\end{array}
\end{equation}
Now, we have
\begin{equation}\label{product5}
\begin{array}[c]{c}
F(q,p) = \int dze^{i pz/\hbar} <q-\frac{z}{2}|\hat{A} \hat{B}|q+\frac{z}{2}>\\ \\
= \int dz dq' e^{ipz/\hbar}<q-\frac{z}{2}|\hat{A}|q'><q'|
\hat{B}|q+\frac{z}{2}>\\ \\ \int dz dq' d\sigma d\sigma ' e^{(i/2\hbar)\sigma
(q'+q-\frac{z}{2})}e^{(i/2\hbar)\sigma '(q'+q+\frac{z}{2})} \alpha(\sigma ,q'-q+\frac{z}{2}) \alpha' (\sigma
',q-q'+\frac{z}{2})e^{ipz/\hbar}.
\end{array}
\end{equation}
By defining the new variables $\tau=q'-q+\frac{z}{2}$ and $\tau
'=q-q'+\frac{z}{2}$, we would get
\begin{equation}\label{product6}
\begin{array}[c]{c}
F(q,p) = \int d \tau d \tau ' d \sigma d \sigma '
\alpha(\sigma ,\tau) e^{(i/\hbar)(\sigma q+ \tau
p)}e^{(i/2\hbar)(\sigma '\tau - \sigma \tau ')}\\ \\ \times
e^{(i/\hbar)(\sigma ' q+ \tau ' p)} \alpha' (\sigma ',\tau ').
\end{array}
\end{equation}
Now, consider the exponential between the other two exponentials
and Taylor expand it. Consider the second term while forget about
all constants, i.e., $\sigma ' \tau - \sigma \tau '$, it is easy to
see that by differentiation of the exponential on the right with
respect to $p$ and the exponential on the left with respect to
$q$, we can get $\sigma ' \tau$. We can get $\sigma \tau '$ by
differentiation of the exponential on the right with respect to
$q$ and the exponential on the left with respect to $p$.
Therefore, replacement of $(i/\hbar)(\sigma ' \tau - \sigma \tau
')$ by $(\hbar \Lambda /2i)$ makes no difference up to the second
term in the Taylor expansion, by more elaboration you can show
that this is also true for the higher order terms. After the
mentioned replacement $A(p,q)$ and $B(p,q)$ (inverse of (\ref{weyl1}))
will appear in (\ref{product6}), and we will get the first
equality in (\ref{product1}). If we change the place of the first
two and the last two terms in (\ref{product6}), we can repeat the
preceding discussion by interchanging the differentiation with
respect to $p$ by the differentiation with respect to $q$ and vice
versa. Thus, we can easily get the second equality in
(\ref{product1}).

Another way of writing the product of two
operators are Bopp operators which are defined as \cite{kubo}
\begin{equation}\label{product7}
Q=q-\frac{\hbar}{2i}\frac{\partial}{\partial p},\hspace{2cm}
P=p+\frac{\hbar}{2i}\frac{\partial}{\partial q}.
\end{equation}
By taking a test function $f$ and a little elaboration you can
show that $[\sigma q+\tau p, \tau \frac{\partial}{\partial
q}-\sigma \frac{\partial}{\partial p}]=0$. This equality yields
\begin{equation}\label{product8}
exp\left\{\frac{i}{\hbar}[\sigma(q-\frac{\hbar}{2i}\frac{\partial}{\partial
p})+\tau (p+\frac{\hbar}{2i}\frac{\partial}{\partial
q})]\right\}=e^{\frac{i}{\hbar}(\sigma q+ \tau
p)}e^{\frac{1}{2}(\tau \frac{\partial}{\partial q}- \sigma
\frac{\partial}{\partial p})}.
\end{equation}
If we multiply both sides by $e^{(\frac{i}{\hbar})(\sigma 'q+\tau
'p)}$, Taylor expand the middle term on the right hand side,
 and operate it on the exponential on its right, then every $\frac{\partial}{\partial p}$ will be replaced by $\tau '$
and every $\frac{\partial}{\partial q}$ will be replaced by
$\sigma '$. Then, we will have the Taylor expansion of an
exponential function in the middle, which can be gathered and give
the final relation
\begin{equation}\label{product9}
\begin{array}[c]{c}
exp\left\{\frac{i}{\hbar}[\sigma(q-\frac{\hbar}{2i}\frac{\partial}{\partial
p})+\tau (p+\frac{\hbar}{2i}\frac{\partial}{\partial q})]\right\}
e^{\frac{i}{\hbar}(\sigma 'q+\tau 'p)}\\ \\
=e^{\frac{i}{\hbar}(\sigma q+ \tau p)}e^{\frac{i}{2\hbar}(\tau
\sigma '- \sigma \tau ')}e^{\frac{i}{\hbar}(\sigma 'q+\tau 'p)}.
\end{array}
\end{equation}
 On the right hand side of (\ref{product9}), we have all the exponential
terms we had on the right hand side of (\ref{product6}). By
replacing them and using the notation introduced in
(\ref{product7}), we get
\begin{equation}\label{product10}
\begin{array}[c]{c}
F(q,p) = \int d \tau d \tau ' d\sigma d \sigma '
\alpha(\sigma ,\tau) e^{\frac{i}{\hbar}(\sigma Q+ \tau
P)}e^{\frac{i}{\hbar}(\sigma 'q + \tau 'p)} \alpha'(\sigma ',\tau
').
\end{array}
\end{equation}
Now, we can define
\begin{equation}\label{product11}
\tilde{A}(Q,P) \equiv \int d \tau d \sigma \alpha(\sigma
,\tau) e^{\frac{i}{\hbar}(\sigma Q+ \tau P)}.
\end{equation}
Therefore, we can express $F(p,q)$ as
\begin{equation}\label{product12}
F(p,q)=\tilde{A}(Q,P) B(p,q).
\end{equation}
Similarly it can be shown that
\begin{equation}\label{product13}
F(p,q)=\tilde{B}(Q^*,P^*) A(p,q),
\end{equation}
where
\begin{equation}\label{product14}
Q^*=q+\frac{\hbar}{2i}\frac{\partial}{\partial p},\hspace{2cm}
P^*=p-\frac{\hbar}{2i}\frac{\partial}{\partial q}.
\end{equation}
We know that the Wigner function is the function associated with
$(\frac{1}{2 \pi \hbar})^n \hat{\rho}$, and that the equation of
motion for $\rho$ is
\begin{equation}\label{product15}
i \hbar \partial{\hat{\rho}}/\partial t=[\hat{H},\hat{\rho}].
\end{equation}
Using the product rule, we can transform
(\ref{product15}) to
\begin{equation}\label{product16}
i \hbar \partial P/\partial t=H(q,p)e^{\hbar
\Lambda/2i}P(q,p)-P(q,p)e^{\hbar \Lambda/2i}H(q,p).
\end{equation}
The first term of the Taylor expansion is $HP-PH$, which is equal
to zero. The second term in the Taylor expansion of the first and
the second term of (\ref{product16}) are just negative of each
other so they build up to $\frac{\hbar}{i}[\frac{\partial
H}{\partial p}\frac{\partial P}{\partial q}-\frac{\partial
H}{\partial q}\frac{\partial P}{\partial p}]$. For the third term
we need
\begin{equation}\label{product17}
\Lambda^2=(\frac{\overleftarrow{\partial}}{\partial
p}\frac{\overrightarrow \partial}{\partial
q}-\frac{\overleftarrow{\partial}}{\partial
q}\frac{\overrightarrow
\partial}{\partial p})(\frac{\overleftarrow{\partial}}{\partial
p}\frac{\overrightarrow \partial}{\partial
q}-\frac{\overleftarrow{\partial}}{\partial
q}\frac{\overrightarrow
\partial}{\partial p})
\end{equation}
 By inserting two test functions $f$ and $g$, respectively, in the left and right hand side of the expression in (\ref{product17}), we can show that
\begin{equation}\label{product18}
\Lambda^2=\frac{\overleftarrow{\partial^2}}{\partial
p^2}\frac{\overrightarrow{\partial^2}}{\partial
q^2}-2\frac{\overleftarrow{\partial^2}}{\partial q \partial
p}\frac{\overrightarrow{\partial^2}}{\partial p \partial
q}+\frac{\overleftarrow{\partial^2}}{\partial
q^2}\frac{\overrightarrow{\partial^2}}{\partial p^2}
\end{equation}
In order to generalize the expression for $\Lambda^2$ to higher
dimensions, we can write
\begin{equation}\label{product22}
\Lambda^2=\left[\sum_i(\frac{\overleftarrow{\partial}}{\partial
p_i}\frac{\overrightarrow \partial}{\partial
q_i}-\frac{\overleftarrow{\partial}}{\partial
q_i}\frac{\overrightarrow
\partial}{\partial p_i})\right]\left[\sum_j(\frac{\overleftarrow{\partial}}{\partial
p_j}\frac{\overrightarrow \partial}{\partial
q_j}-\frac{\overleftarrow{\partial}}{\partial
q_j}\frac{\overrightarrow
\partial}{\partial p_j})\right],
\end{equation}
in order to get
\begin{equation}\label{product23}
\Lambda^2=\sum_{i,j}\left[\frac{\overleftarrow{\partial^2}}{\partial
p_i \partial p_j}\frac{\overrightarrow{\partial^2}}{\partial q_i
\partial q_j}-2\frac{\overleftarrow{\partial^2}}{\partial q_i
\partial p_j}\frac{\overrightarrow{\partial^2}}{\partial p_i
\partial q_j}+\frac{\overleftarrow{\partial^2}}{\partial q_i
\partial q_j}\frac{\overrightarrow{\partial^2}}{\partial p_i
\partial p_j}\right].
\end{equation}
Because of the symmetry of (\ref{product18}), the third term in
(\ref{product15}) is zero. Again, by inserting the test functions
$f$ and $g$, we can evaluate
\begin{equation}\label{product19}
\begin{array}[c]{c}
\Lambda^3=\Lambda^2
\Lambda=(\frac{\overleftarrow{\partial^2}}{\partial
p^2}\frac{\overrightarrow{\partial^2}}{\partial
q^2}-2\frac{\overleftarrow{\partial^2}}{\partial q \partial
p}\frac{\overrightarrow{\partial^2}}{\partial p \partial
q}+\frac{\overleftarrow{\partial^2}}{\partial
q^2}\frac{\overrightarrow{\partial^2}}{\partial
p^2})(\frac{\overleftarrow{\partial}}{\partial
p}\frac{\overrightarrow \partial}{\partial
q}-\frac{\overleftarrow{\partial}}{\partial
q}\frac{\overrightarrow
\partial}{\partial p})\\ \\ =\left(\frac{\overleftarrow{\partial^3}}{\partial
p^3}\frac{\overrightarrow{\partial^3}}{\partial
q^3}-3\frac{\overleftarrow{\partial^3}}{\partial q \partial
p^2}\frac{\overrightarrow{\partial^3}}{\partial p \partial
q^2}+3\frac{\overleftarrow{\partial^3}}{\partial q^2 \partial
p}\frac{\overrightarrow{\partial^3}}{\partial p^2 \partial
q}-\frac{\overleftarrow{\partial^3}}{\partial
q^3}\frac{\overrightarrow{\partial^3}}{\partial p^3}\right),
\end{array}
\end{equation}
and
\begin{equation}\label{product20}
\begin{array}[c]{c}
\Lambda^4=\Lambda^3
\Lambda=\left(\frac{\overleftarrow{\partial^3}}{\partial
p^3}\frac{\overrightarrow{\partial^3}}{\partial
q^3}-3\frac{\overleftarrow{\partial^3}}{\partial q \partial
p^2}\frac{\overrightarrow{\partial^3}}{\partial p \partial
q^2}+3\frac{\overleftarrow{\partial^3}}{\partial q^2 \partial
p}\frac{\overrightarrow{\partial^3}}{\partial p^2 \partial
q}-\frac{\overleftarrow{\partial^3}}{\partial
q^3}\frac{\overrightarrow{\partial^3}}{\partial
p^3}\right)(\frac{\overleftarrow{\partial}}{\partial
p}\frac{\overrightarrow \partial}{\partial
q}-\frac{\overleftarrow{\partial}}{\partial
q}\frac{\overrightarrow
\partial}{\partial p})\\ \\ =\left(\frac{\overleftarrow{\partial^4}}{\partial
p^4}\frac{\overrightarrow{\partial^4}}{\partial
q^4}-4\frac{\overleftarrow{\partial^4}}{\partial q \partial
p^3}\frac{\overrightarrow{\partial^4}}{\partial p \partial
q^3}+6\frac{\overleftarrow{\partial^4}}{\partial q^2 \partial
p^2}\frac{\overrightarrow{\partial^4}}{\partial p^2 \partial
q^2}-4\frac{\overleftarrow{\partial^4}}{\partial q^3 \partial
p}\frac{\overrightarrow{\partial^4}}{\partial p^3 \partial
q}+\frac{\overleftarrow{\partial^4}}{\partial
q^4}\frac{\overrightarrow{\partial^4}}{\partial p^4}\right).
\end{array}
\end{equation}
By continuing in this manner we can show that
\begin{equation}\label{product21}
\hbar \partial P/\partial t=-2H(q,p)sin(\hbar \Lambda/2)P(q,p).
\end{equation}

\section{Proof of the impossibility of a positive phase space probability distribution}
Wigner was aware that this probability function gets
negative values (unless the world was just made up of Gaussian
wave packets). Thus, he emphasized that this is just a
calculational tool, not a real probability distribution in phase
space. Latter, he uses the fact that for a mixed
state $P(q,p)=\sum w_iP_i(q,p)$, where $w_i$ is the
probability of the i'th pure state, and $P_i(q,p)$ is the Wigner
function for the i'th pure state. Then, he shows that by imposing
the conditions (i) and (ii) (see section (\ref{Properties})) it is
impossible to build an always positive distribution function \cite{lande}. He
used $\psi(q)=a\psi_1(q)+b\psi_2(q)$, where $\psi_1$ is zero
outside $I_1$ and $\psi_2$ is zero outside of $I_2$. Consider
$I_1$ and $I_2$ to be two non-overlapping intervals over the space
of coordinate. Now we have
\begin{equation}\label{negativeness1}
P_{ab}(q,p)=|a|^2P_1+a^*bP_{12}+ab^*P_{21}+|b|^2P_2.
\end{equation}
 If $q$ is outside of $I_1$, $P_1$ is zero for such
a $q$ and the only way to have a positive value for $P_{ab}(q,p)$
for every $a$ and $b$ is to have $P_{12}(q,p)=P_{21}(q,p)=0$. The
same reasoning can be given for the $q$ outside of $I_2$.
Therefore, every where, we have
\begin{equation}\label{negativeness2}
P_{ab}(q,p)=|a|^2P_1+|b|^2P_2.
\end{equation}
This means that $P_{ab}$ is independent of the complex phase of
$a/b$. Consider the Fourier transform
of $\psi_1$ and $\psi_2$ to be $\phi_1(p)$ and $\phi_2(p)$. By
removing $P_{ab}$ from equation (\ref{negativeness1}) and equation
(\ref{negativeness2}), then integrating both sides of the
resultant equation with respect to $q$ and using the mentioned
Fourier inverses, we get
\begin{equation}\label{negativeness3}
\begin{array}[c]{c}
|a|^2\int P_1(q,p)dq+|b|^2\int P_2(q,p)dq \\ \\
=|a|^2|\phi_1(p)|^2+|b|^2|\phi_2(p)|^2+2 Re [ab^*
\phi_1(p)\phi_2(p)^*].
\end{array}
\end{equation}
For this relation to be valid for all $a$ and $b$, we must have
\begin{equation}\label{negativeness4}
\phi_1(p)\phi_2(p)^*=0.
\end{equation}
On the other hand, $\phi_1$ and $\phi_2$ are Fourier transforms of
confined functions; thus, they cannot vanish on a finite
interval. This is a contradiction, and QED. Because, it seems
possible to break down every normalizable wave function into such
a linear combination, thus this proof excludes the possibility of
having a phase space distribution for a quantum state. Wigner
\cite{lande}, also, showed that by imposing the conditions (i)-(v)
(\ref{wigner1}) is unique. While, O'connell and Wigner
\cite{oconnell} show that by imposing conditions (i)-(iv) and (vi)
(\ref{wigner1}) is the only possible distribution.

\section{Dynamics of the Wigner function}
If we want to express quantum mechanics in terms of the Wigner
function we must derive Wigner functions equation of motion. This
will be done with the aid of the Schroedinger equation, i.e.,
\begin{equation}\label{schroedinger}
i \hbar \frac{\partial{\psi(t)}}{\partial{t}} =
\left[-\frac{\hbar^2}{2m}\frac{\partial^2}{\partial
q^2}+V(q,t)\right] \psi(t).
\end{equation}
By conjugate transposing both sides of the Schroedinger equation,
we get
\begin{equation}\label{schroedinger*}
-i \hbar \frac{\partial{\psi(t)^*}}{\partial{t}} =\left
[-\frac{\hbar^2}{2m}\frac{\partial^2}{\partial q^2}+V(q,t)\right]
\psi(t)^*.
\end{equation}
 Decomposing the time dependence of
$P$ into two parts, we have
\begin{equation}\label{dynamic1}
\begin{array}[c]{c}
\frac{\partial{P}}{\partial{t}} = (\frac{1}{\pi \hbar})^n \int dy
\left[\frac{\partial{\psi(q+y)^*}}{\partial{t}}\psi(q-y)+\psi(q+y)^*
\frac{\partial{\psi(q-y)}}{\partial{t}}\right]e^{2ipy/\hbar} \\ \\
=\frac{\partial_k P}{\partial{t}}+\frac{\partial_v
P}{\partial{t}}.
\end{array}
\end{equation}
In the last expression of (\ref{dynamic1}), the first part arises
from the kinetic part of the Hamiltonian and the second part arises
from its potential part. By substituting (\ref{schroedinger}) and
(\ref{schroedinger*}) in (\ref{dynamic1}) and considering the n to
be equal to one, we can get
\begin{equation}\label{dynamic3}
\frac{\partial_k{P}}{\partial{t}}=(\frac{-i}{2\pi m}) \int dy
\left[\frac{\partial^2{\psi(q+y)^*}}{\partial{y^2}}\psi(q-y)-\psi(q+y)^*
\frac{\partial^2{\psi(q-y)}}{\partial{y^2}}\right]e^{2ipy/\hbar},
\end{equation}
where we have replaced $\partial^2/\partial q^2$ by
$\partial^2/\partial y^2$. Integration by parts, because $\psi$
vanishes at $-\infty$ and $\infty$, yeilds
\begin{equation}\label{dynamic4}
\frac{\partial_k{P}}{\partial{t}}=(\frac{-p}{\pi \hbar m}) \int dy
\left[\frac{\partial{\psi(q+y)^*}}{\partial{y}}\psi(q-y)-\psi(q+y)^*
\frac{\partial{\psi(q-y)}}{\partial{y}}\right]e^{2ipy/\hbar}.
\end{equation} By going back to $\partial/\partial{q}$, we obtain
\begin{equation}\label{dynamic5}
\frac{\partial_k{P}}{\partial{t}}=-\frac{p}{m}\frac{\partial{P(q,p)}}{\partial
q}, \end{equation} which is identical to the corresponding term inthe classical Liouville equation. Also, we have
\begin{equation}\label{dynamic6}
\begin{array}[c]{c}
\frac{\partial_v{P}}{\partial{t}}=\frac{i}{(\pi \hbar)^n \hbar}
\int dy [(V\psi)(q+y)^*\psi(q-y)-\psi(q+y)^*(V\psi)(q-y)]e^{2ipy/\hbar}\\
\\=\frac{i}{(\pi \hbar)^n \hbar} \int dy
[V(q+y)-V(q-y)]\psi(q+y)^*\psi(q-y)e^{2ipy/\hbar}.
\end{array}
\end{equation}
By Taylor expanding $V$, we get
\begin{equation}\label{dynamic7}
V(q+y)=\sum_{\lambda=0}^\infty \frac{y^\lambda}{\lambda
!}\frac{\partial^\lambda V}{\partial q^\lambda}.
\end{equation}
Therefore, we have
\begin{equation}\label{dynamic8}
\frac{\partial_v{P}}{\partial{t}}=\frac{2i}{\pi \hbar^2}\int
dy\sum_\lambda \frac{y^\lambda}{\lambda !}\frac{\partial^\lambda
V}{\partial q^\lambda}\psi(q+y)^*\psi(q-y)e^{2ipy/\hbar},
\end{equation}
where the sum is over the odd positive integers $\lambda$, since
the even terms resulting from $V(q+y)$ and those resulting from
$V(q-y)$ cancel each other. Because by differentiating the
exponential term with respect to $p$, we get a $y$ multiplier,
$y^\lambda$ can be replaced with $[(\hbar/2i)(\partial /\partial
p)]^\lambda$ to get
\begin{equation}\label{dynamic9}
\frac{\partial_v{P}}{\partial{t}}=\sum_\lambda \frac{1}{\lambda
!}(\frac{\hbar}{2i})^{\lambda-1}\frac{\partial^\lambda
V(q)}{\partial q^\lambda}\frac{\partial^\lambda {P(q,p)}}{\partial
p^\lambda}.
\end{equation}
For the sake of simplicity, equations
(\ref{dynamic7})-(\ref{dynamic9}) are written for the one
dimensional case ($n=1$). In order to generalize them to higher
dimensions $\lambda !$ should be replaced by $\Pi_i \lambda_i !$,
any thing to power $\lambda$ with the same thing to power
$\sum_i \lambda_i$, $\partial q^\lambda$ with $\Pi_i \partial
q^\lambda_i$, and $\partial p^\lambda$ with $\Pi_i \partial
p_i^{\lambda_i}$. But, remember
$\lambda_i$'s take positive integers that yield an odd positive
integer for $\sum_i \lambda_i$. We can, also, write
$\frac{\partial_v{P}}{\partial{t}}$ in the form
\begin{equation}\label{dynamic10}
\frac{\partial_v{P}}{\partial{t}}=\int dj P(q,p+j)J(q,j),
\end{equation}
while
\begin{equation}\label{dynamic11}
\begin{array}[c]{c} 
J(q,j)=\frac{i}{(\pi \hbar)^n\hbar}\int dy
[V(q+y)-V(q-y)]e^{-2ijy/\hbar}\\ \\=\frac{1}{(\pi
\hbar)^n\hbar}\int dy [V(q+y)-V(q-y)]sin(2jy/\hbar)
\end{array}
\end{equation}
has interpreted as the probability of a jump in momentum by an
amount $j$, if the position is $q$. We can go from the first to the
second equality in (\ref{dynamic11}), because
$e^{ix}=cos(x)+isin(x)$ and the function in the square brackets is
an odd function so when it is multiplied by an even function,
$cos(2jy/\hbar)$ and integrated over the whole space, it will give
zero. 

Now, we are able to get the equation of motion as
\begin{equation}\label{dynamic12}
\frac{\partial P}{\partial
t}=-\sum_{i=1}^n\frac{p_i}{m_i}\frac{\partial P}{\partial
q_i}+\sum\frac{\partial^{\sum \lambda_i}V}{\Pi_i \partial
q_i^{\lambda_i}}\frac{(\hbar/2i)^{\sum \lambda_i-1}}{\Pi \lambda_i
!}\frac{\partial^{\sum \lambda_i}P}{\Pi \partial p_i^{\lambda_i}}.
\end{equation}
Now consider the case where the potential has no third or higher
order derivative; then, evidently (\ref{dynamic11}) is the
classical Liouville equation. For a system consisting of a bunch
of harmonic oscillators and free particles, surprisingly we can solve the
easier classical equations of motion and get the exact quantum
result!

\section{Attempts for giving a
probability distribution interpretation to the Wigner function}
Some people have argued this function as a valid probability
distribution, and some others argue it as a valid probability
distribution just for some situations. Stenholm has argued that,
we can "obtain verifiable predictions" only by using "suitable
test bodies." He show that always positive probabilities come out
of the Wigner distribution when these arguments are implemented.
In relativistic quantum mechanics \cite{barut}, even if we are
working in the position representation, the choice of the position
observable is not at all trivial. Therefore, probably in
non-relativistic cases it is just the absence of mathematical
complications which make us to believe that we can make a
classical interpretation of position and momentum.

The way to measure the momentum and coordinate of a particle is to
let it interact with another body which we usually let approach
it's classical limit. This second body is a test particle, and we
are actually performing a scattering experiment. In this
scattering experiment, test particle transfers the desired
information out of the interaction region. "Only probability
distributions observable in this manner can be given a physical
interpretation" \cite{stenholm}. In cases where the test particle
is carrying both coordinate and momenta information, restrictions
due to Heisenberg uncertainty principle should be taken into
account. Stenholm emphasized that, in order to get the most
precise results for both momentum and coordinate, we must use a
test particle which is in a state of minimum uncertainty. At the
end, the test body is bringing out some information which at best
allow us to confine our system to a region of phase space
satisfying the relation $\Delta p \Delta q \leq \frac{\hbar}{2}$
and no more precise detail is achievable. The minimum uncertainty
wave packet can be determined uniquely as
\begin{equation}\label{interpretation1}
\psi_0(x)=Cexp\left(-\frac{(x-<x>)^2}{4b^2}+\frac{i<P>x}{\hbar}\right),
\end{equation}
where $<>$ denotes the expectation value, and we have
uncertainties $\Delta x =b$ and $\Delta p=\hbar/2b$. The Wigner
function for this minimum uncertainty wave packet is
\begin{equation}\label{interpretation2}
W_0(R,P)=Aexp\left(-\frac{(R-<x>)^2}{2b^2}-\frac{4b^2(P-<P>)^2}{2\hbar^2}\right).
\end{equation}
Stenholm argued that "a Wigner function $W(P,R)$ is not directly
observable but has to be convoluted with the function describing
the test particle, which smears it, at least, by the amount
implied by the function" (\ref{interpretation2}). This convolution
leads to
\begin{equation}\label{interpretation3}
\begin{array}[c]{c}
P(\pi,q)=A\int \int
exp\left(-\frac{(R-q)^2}{2b^2}-\frac{2b^2(P-\pi)^2}{\hbar^2}\right)W(R,P)dRdP
\\ \\ =(\frac{A}{\pi \hbar})\int \int \int drdPdR
e^{(-iPr/\hbar)}e^{-(R-q)^2/2b^2}e^{-2b^2(P-\pi)^2/\hbar^2}\psi(R+\frac{r}{2})\psi(R-\frac{r}{2})^*.
\end{array}
 \end{equation}
Here $(q,\pi)$ are a couple of position and momentum coordinate,
and Stenholm hope them to give meaningful phase space
interpretation. $A$ is just a normalization constant. By carrying
out the integral over $P$, we can show that $P(q,\pi)\geq0$. The
immediate criticism to this approach is that not all test bodies
are minimum uncertainty wave packets. Thus, Stenholm emphasize
that in order for the test body to exhibit a nearly classical
behavior it must be smooth and every wave function smoother than
the minimum uncertainty wave packet will fulfill the positiveness
requirement. All in all, we have to calculate $W(P,R;t)$ up to the
moment of measurement, and then smooth it to obtain $P(\pi,q;t)$.
Since, $P(\pi,q;t)$ depends on the test particle prepared by the
observer it has no dynamics. The idea of smoothing the Wigner
function with a Gaussian function was first introduced by Husimi
\cite{husimi}. He get the positive distribution which is now
called the Husimi distribution. Husimi didn't interpret it as a
phase space distribution because it doesn't poses the property
(ii). There is also a bunch of work for interpreting Quantum
Optics based on Wigner function. Including Marshall and Santos
\cite{marshall}, who argue that, there is just a subset of states
in the Hilbert space which can be generated in the laboratory, and
those states "May be represented by a positive Wigner
distribution." They have also claimed that \cite{marshall2}, the
experiments which are exhibiting non-classical behavior of light
can be interpreted just by assuming light as an electromagnetic
wave in accordance with the Maxwell's equations of motion. Also,
Holland \emph{et al} \cite{holland} published on the "Relativistic
generalization of the Wigner function and its interpretation in
the causal stochastic formulation of quantum mechanics." There are
some good reviews on the mathematical properties and applications
of the Wigner function, e.g., \cite{hillery}, \cite{imre} and
\cite{berry}.
 \section{Discussion} We can argue that, any
experiment which is designed to measure both momentum and
coordinate will get average information about a region of phase
space which is large enough to give a positive value. The Wigner
function give a positive average over this region and thus
interpreting the Wigner function as a probability distribution
function is experimentally adequate. But does it make sense to
consider a probability function which has no meaning on a point
but is representing a physical reality when averaged over any
large enough interval! Thats weird but is it more weird than a
particle owning either momentum or position and not both of them
at the same time? The classical formalism of quantum mechanics
doesn't provide any prediction regarding a simultaneous
measurement of position and momentum. If it was possible to
perform such an experiment, for example, as suggested, thorough
sending out a test particle to interact with system (quantum
particle) and be able to measure its position and momentum before
and after the interaction, because it is approaching classical
behavior, then we could perform an experimental test for this
interpretation. But it seems that no semi-classical test particle
can be used for a measurement on a quantum system without
enormously changing its state.
\bibliography{bibliographytermpaper}
\end{document}